\documentclass[twocolumn,amsmath,prl,showpacs]{revtex4}

\input{epsf}
\epsfclipon

\begin{document}

\title{Zero-temperature criticality in the two-dimensional gauge glass model}

\author{Lei-Han Tang$^1$}\email{lhtang@hkbu.edu.hk} 
\author{Peiqing Tong$^{1,2}$}
\affiliation{$^1$Department of Physics, Hong Kong Baptist University, Kowloon Tong, Hong Kong SAR, China\\
$^2$Department of Physics, Nanjing Normal University, Nanjing, Jiangsu 210097, China
}

\date{\today}

\begin{abstract}
The zero-temperature critical state of the two-dimensional gauge glass model 
is investigated. It is found that low-energy vortex configurations afford a simple
description in terms of gapless, weakly interacting vortex-antivortex pair 
excitations. A linear dielectric screening calculation is presented
in a renormalization group setting that yields a power-law decay of 
spin-wave stiffness with distance. These properties are in agreement 
with low-temperature specific heat and spin-glass susceptibility data
obtained in large-scale multi-canonical Monte Carlo simulations. 

\end{abstract}

\maketitle

Rigidity is a key concept that unifies description of long-ranged
order in magnets, superfluids, superconductors, and crystalline solids.
\cite{Anderson} For two-dimensional (2D) systems with an O(2) symmetry, 
the helicity modulus provides a quantitative measure of rigidity,
and assumes a finite value in a pure sample at sufficiently low 
temperatures.\cite{FBJ,NK77}
There are however a number of experimental situations, notably
granular thin films of high-$T_c$ superconductors, where one has
to deal with quenched disorder that introduces random frustration
and ground state vortices into the system.\cite{Dekker,Fisher89,newrock} 
Understanding how rigidity weakens or disappears altogether is crucial 
for interpreting the often complex equilibrium and dynamic behavior of 
these systems at low temperatures.\cite{Hyman}

The gauge glass model\cite{Ebner,FTY} in two dimensions offers 
a good example where the nature of complexity and glassiness associated 
with strong disorder can be examined in quantitative detail.
Two competing scenarios regarding the low-temperature phase diagram
have been proposed:
i) The system is disordered at any temperature $T>0$, but critical
at $T=0$, as characterized by a power-law diverging glass correlation length
$\xi_{\rm G}\simeq T^{-\nu}$ with a rather large exponent $\nu\simeq 2.5$.
\cite{FTY,RY93,Akino,Katzgraber,NW04}
ii) There is a glass phase of frozen vortices that undergoes
a continuous transition at $T_c\simeq 0.2J$.\cite{CP99,Kim,HMK}
Direct measurement of rigidity in terms of spin-glass susceptibility
has not been able to differentiate the two scenarios unambiguously
due to the relatively small system sizes examined in simulations.
The domain-wall energy analysis\cite{FTY} has provided some insight on
the characteristics of low-lying excitations, although their true
identity remains mysterious.

In this paper, we describe results of extensive numerical investigation
of the 2D gauge glass model and the related Coulomb gas problem.
We show that, despite strong correlation in vortex positions 
and the collective nature of the ground state,
low energy excited states of this model take the form of a dilute gas
of fermion-like vortex-antivortex pairs which have a gapless, 
excitation spectrum. Based on the existence of such
elementary excitations, we present a phenomenological
renormalization group (RG) calculation that explains the power-law decay of 
spin-wave stiffness $J(R)\sim R^\theta$ with distance $R$ in the ground state.
Evidence for gapless pair excitations is seen in the linear
specific heat at low temperatures. The stiffness exponent $\theta$
is estimated from the strength of the screened disorder potential in
the ground state of the Coulomb gas problem, and from the
spin-glass susceptibility data obtained in large-scale multicanonical
Monte Carlo simulations of the gauge glass model.
In both cases, we obtain $\theta\simeq -0.45$,
in approximate agreement with earlier studies.

The gauge glass model is defined by the Hamiltonian,
\begin{equation}
H=-J\sum_{\langle ij\rangle}\cos(\phi_i-\phi_j-A_{ij}),
\label{gauge-glass-H}
\end{equation}
where $\phi_i$ is the phase variable on site $i$, $J$ is the 
coupling constant, and $A_{ij}$ are quenched random variables
uniformly distributed on $[-\pi,\pi)$. Summation is over neareast
neighbor pairs of sites, here on a square lattice. Decomposing
the lattice gradient of the phase field $\phi_i$ into rotation-free 
and divergence-free components, one arrives at the following Coulomb gas
model of vortices,\cite{rubi83}
\begin{equation}
H_v=\sum_i(m_i^2E_c+m_iV_i)-2\pi J\sum_{i<j}m_i m_j\ln {r_{ij}\over a}.
\label{coulomb-gas}
\end{equation}
Here $m_i$ is the vortex charge on site $i$, $E_c$ is the vortex core energy, 
$a$ is the lattice constant, and $r_{ij}$ is the distance between sites 
$i$ and $j$. The random potential $V_i$ can be expressed as a sum of dipolar
potentials produced by the random phase shifts $A_{ij}$. 
Its statistics is specified by a logarithmically growing variance 
$\langle V_i^2\rangle\simeq 2\pi\sigma J^2\ln (L/a)$, and 
logarithmic spatial correlations 
$\langle (V_i-V_j)^2\rangle\simeq 4\pi\sigma J^2\ln(r_{ij}/a)$.
For the gauge glass model, we have $\sigma=\pi^2/3$ and $E_c\simeq 5J$, such
that approximately one third of the sites are occupied by either a vortex or 
an antivortex in the ground state.

Due to the high density of vortices and antivortices, the 
RG analysis\cite{nskl,tang96} developed for a dilute Coulomb gas 
can not be applied here. We have investigated numerically 
the minimal energy states of the Coulomb gas Hamiltonian (\ref{coulomb-gas}) 
using a greedy algorithm, restricting vortex charges to $m_i=0,\pm 1$.
Such states are constructed through successive addition and removal of 
vortex-antivortex pairs, each time picking a pair state of the lowest 
energy. (Note that moving a vortex to a vacant site can also be realized by 
such a move.) The process terminates when the system energy can not be 
lowered further through single pair addition or removal. In an improved
ground state search, we allow the process to continue
until a previous vortex configuration is revisited.
Our experience shows that, for systems up to a linear size $L=16$, the 
algorithm is able to find the ground state typically in 100 or fewer trials
of random initial conditions.

Figure 1 illustrates results of our finding for a $16\times 16$ system.
The ground state vortex (solid circle) and antivortex (open circle) 
configuration $\{ m_i^0\}$ is shown in Fig. 1(a).
As expected, there is no apparent order of the vortices
and antivortices. The same is true also for other low energy states
obtained under the greedy algorithm. However, when the excess charge 
$\{\tilde m_i\equiv m_i-m_i^0\}$ of these states is plotted,
as shown in Figs. 1(b)-(d), we see clearly that the low energy states
overlap strongly with the ground state. The excess charges can be
identified as elementary excitations that include
single vortex/antivortex movement to a new position (e.g., vortex at site A 
to site B, vortex at site C to a neighboring site, antivortex from site G to 
a neighboring site, and antivortex from site H to site I),
and insertion (pair D) or deletions (pairs E and F) of
vortex-antivortex pairs.

\begin{figure}
\epsfxsize=\linewidth
\epsfbox{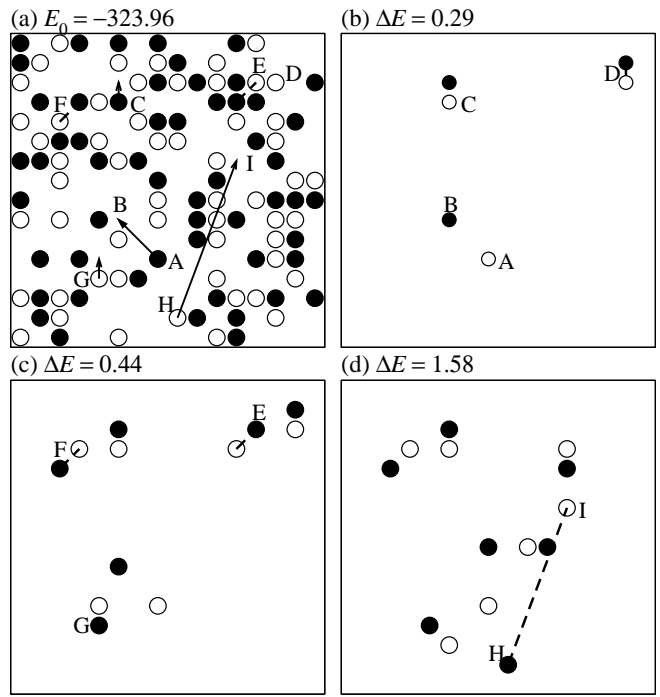}
\caption{(a) Vortices (solid circles) and anti-vortices (open circles) 
in the ground state. (b)-(d) Excited states shown in terms of excess
vortices (solid circles) and antivortices (open circles) from the ground
state. Energies are in units of $J$.
}
\label{vortex_conf}
\end{figure}

Repeated numerical experiments of this type led to the
following general observations:
i) The ground state contains a high density of vortex-antivortex 
pairs of all sizes. 
ii) Low energy excited states can be described as a dilute gas of excess 
vortices from the ground state. 
iii) Most of the excess vortices in ii) form closely bound
vortex-antivortex pairs. However, pairs of larger size 
[e.g., the vortex-antivortex pair at H and I in Fig. 2(d)] 
may also be present. They usually form a complex with smaller pairs that 
provide screening of the Coulomb interaction. 

To characterize the energetics of low-energy pair excitations,
we rewrite Eq. (\ref{coulomb-gas}) 
in terms of the excess vortex charge $\tilde m_i=m_i-m_i^0$,
\begin{equation}
H_v=E_0+\sum_i(\tilde m_i^2E_c+\tilde m_i\tilde V_i)-
2\pi J\sum_{i<j}\tilde m_i \tilde m_j\ln {r_{ij}\over a}.
\label{coulomb-gas-renorm}
\end{equation}
Here $E_0$ is the energy of the ground state $\{m_i^0\}$, and
$\tilde V_i$ is an effective potential at site $i$ that includes the 
original disorder potential $V_i$ as well as the potential produced 
by the ground state vortex population $\{m_i^0\}$. 

In the original model, different Fourier components of the disorder
potential are gaussian distributed and statistically independent of
each other. Their variance is given by
$\langle V({\bf k})V(-{\bf k})\rangle = N\sigma(2\pi J)^2G({\bf k})$,
where $N=L^2$ is the total number of sites and
$G({\bf k})=1/[4\sin^2(k_x/2)+4\sin^2(k_y/2)]$ is the lattice 
Green's function. Figure 2(a) shows 
$\langle \tilde V({\bf k})\tilde V(-{\bf k})\rangle/G({\bf k})$
at $k_y=0$ and $k_x=2\pi/L,4\pi/L,\ldots,\pi$, for ground states
obtained under the greedy algorithm and averaged over many disorder 
realizations. All energies are in units of $J$.
Statistical errors are smaller or comparable to symbol size.
It is evident that, unlike the ``bare'' disorder potential $V_i$,
long wavelength components of $\tilde V_i$ are much reduced in
strength by the ground state vortices. In fact, the data can be
fitted well by assigning a {\bf k}-dependent coupling constant
$J(k_x,0)\sim |\sin(k_x/2)|^{-\theta}$, with $\theta=-0.45$, as 
illustrated by the dashed line in the figure.

We have also examined the size-dependence of the gap $\Delta$ 
between the lowest and second lowest values of $\tilde V_i$ in a given
disorder realization. This quantity offers a measure of the spectral
density of $\tilde V_i$ close to its minimum value, which 
governs vortex movement on scale $L$.
Figure 2(b) shows the distribution of $\Delta$ for three different 
system sizes, which is Poisson like. 
The inset shows a scaling plot of $P(\Delta)$, assuming a power-law
width $\Delta_L\sim L^{-0.45}$. A nearly perfect data collapse
is seen. 

\begin{figure}
\epsfxsize=\linewidth
\epsfbox{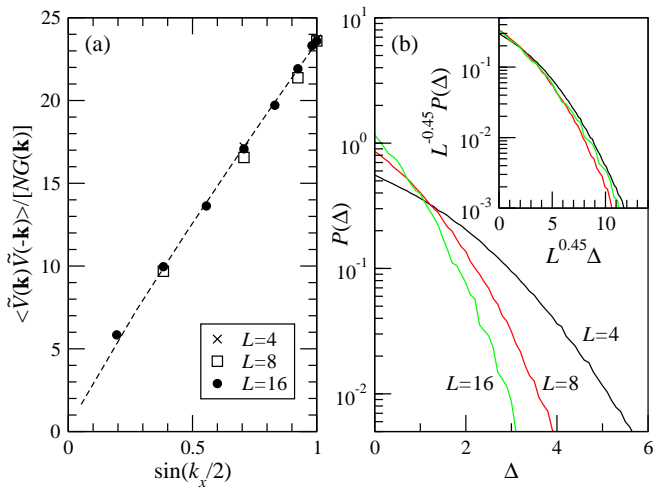}
\caption{(a) Scaled power spectrum of the effective
disorder potential against $k_x$ for three different system sizes. 
The dashed line indicates a power-law fit with an exponent
$2|\theta|=0.9$.
(b) Distribution of the gap energy between the lowest and second lowest
values of the effective potential. Inset shows the scaling plot of the 
distribution with a stiffness exponent $\theta=-0.45$. 
}
\label{v_eff}
\end{figure}

To explain the aforementioned behavior, we consider a dielectric screening
model of vortex-antivortex pairs in a phenomenological RG setting.
Let $R$ be the running length scale in an RG scheme applied to
(\ref{coulomb-gas}), and $J(R)$ be the renormalized interaction strength
after pairs of size less than $R$ have been integrated out.
In an infinite system, the energy $\epsilon$ of 
pair states in the size range $R$ to $R+dR$ forms a continuum that,
for the gauge glass model, extends over both positive and negative values.
Although filling of states with $\epsilon<0$ is a collective process,
the pair-pair Coulomb interaction $V_{\rm pp}\sim r^{-2}$ is not
strong enough to generate a gap at the ``Fermi level''
$\epsilon_{\rm F}=0$.\cite{SE84} Consequently, we expect treating pairs
as independent fermions is qualitatively correct.
An external perturbing field {\bf E}
shifts the energy of state $i$ by an amount $-{\bf E}\cdot {\bf p}_i$,
where ${\bf p}_i={\bf r}^+_i-{\bf r}^-_i$ is the dipole moment of
the pair in state $i$. In the neighborhood of $\epsilon_{\rm F}$,
occupied and unoccupied pair states may switch under the perturbation.
The induced polarization is easily calculated, from which one obtains
the dielectric susceptibility for such a medium at $T=0$,
\begin{equation}
\chi_R=R^2\rho_R(0)\times(dR/R).
\label{chi}
\end{equation}
Here $\rho_R(0)\times(dR/R)$ is the density of pair states per unit area
at $\epsilon=0$. Let $dl=dR/R$, change in $J(R)$ due to this group of
vortex-antivortex pairs can be expressed as\cite{tang96},
\begin{equation}
dJ^{-1}/dl=4\pi^2\hat\rho(0).
\label{J-flow}
\end{equation}
where we have introduced the rescaled density of states 
$\hat\rho(0)\equiv R^2\rho_R(0)$. Since $J(R)$ is the only energy scale at 
zero temperature, we may write, on dimensional grounds,
\begin{equation}
\hat\rho(0)=cJ^{-1},
\label{rho_R}
\end{equation}
where $c$ is some constant.\cite{note}
Combining Eqs. (\ref{J-flow}) and (\ref{rho_R}), we obtain
a power-law decay of the coupling constant
$J(R)=J_{\rm B}R^\theta$ in the ground state, with
$\theta=-4\pi^2c$. At a finite temperature $T$, Eq. (\ref{rho_R})
must be modified to take into account thermally excited pairs
when one reaches a scale $\xi_{\rm G}(T)$ set by $J(\xi_{\rm G})\simeq T$.
This yields a correlation length $\xi_{\rm G}\sim T^{-1/|\theta|}$ beyond
which rigidity disappears. 

We have carried out extensive simulations of the gauge-glass model using
a multi-canonical Monte Carlo sampling scheme that allows one
to equilibrate systems of size up to $L=48$ down to $T=0.05$. 
Computation were performed on a PC cluster using up to 64 nodes. The 
parallel setup is ideal for calculating quantities involving the overlap 
$q\equiv N^{-1}\sum_j\exp[i(\phi_j^a-\phi_j^b)]$ of 
configurations from two different replicas $a$ and $b$ under the same disorder.
Details of our simulation algorithm will be reported elsewhere.

Figure 3 shows the specific heat data for the $L=48$ system, averaged over
26 disorder realizations. Comparison with data at smaller $L$ (not shown)
indicates that no significant finite-size correction is present.
Data on the low temperature side can be fitted to the formula
$c_V=0.5+0.32 T+0.8T^2$, which has a simple physical interpretation.
 The constant term is due to spin-wave contributions.
The linear part confirms presence of gapless 
vortex-antivortex pair excitations in the system.

\begin{figure}
\epsfxsize=8cm
\epsfbox{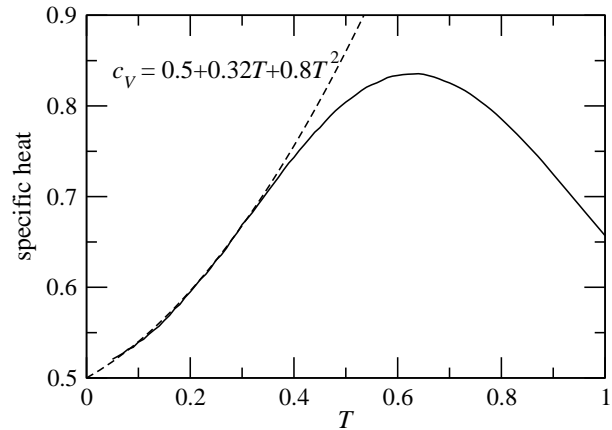}
\caption{Specific heat versus temperature for a $48\times 48$ system.
Dashed line shows a quadratic fit to the low temperature data.}
\label{c_v}
\end{figure}

We have also measured the {\bf k}-dependent spin-glass susceptibility
\begin{equation}
\chi_{\rm SG}({\bf k})=\sum_jC_{\rm SG}(r_{ij})\exp(i{\bf k}\cdot{\bf r}_{ij}),
\label{chi_SG}
\end{equation}
where 
$C_{\rm SG}(r_{ij})=\langle\bigl|\overline{e^{i(\phi_i-\phi_j)}}\bigr|^2\rangle$ 
is the correlation function of the glass order parameter.
Here the overline bar denotes thermal average and $\langle\cdot\rangle$ 
denotes average over the disorder.
Figure 4(a) shows $\chi_{\rm SG}(0)\equiv N\langle\overline{|q|^2}\rangle$
against $T$ for eight different system sizes. 
In agreement with previous studies, the glass phase
correlation grows rapidly at low temperatures. For the largest system
shown at $L=48$, crossover to finite-size dominated regime takes place
already at $T\simeq 0.3$. 

Since spin-wave fluctuation introduces a power-law decay of
$C_{\rm SG}(r)$ at short distances with a temperature-dependent exponent 
$\eta(T)$, extracting $\xi_{\rm G}(T)$ from the finite-size scaling 
ansatz\cite{Katzgraber}
$\chi_{\rm SG}({\bf k},T,L)=L^{2-\eta}\hat\chi(L{\bf k},\xi_{\rm G}/L)$ 
is somewhat ambiguous. Instead, one may consider the ratio
$\alpha(T)\equiv\chi_{\rm SG}(|{\bf k}|=2\pi/L)/\chi_{\rm SG}(0)$ which tends to 1
for $\xi_{\rm G}\ll L$ and 0 for $\xi_{\rm G}\gg L$.
Crossover between the two regimes takes place at $\xi_{\rm G}=L$.
Figure 4(b) shows the derivative $d\alpha(T)/d\ln T$ for six different system
sizes ranging from $L=6$ to 32. The peak position of these curves 
yields a quantitative measure of $\xi_{\rm G}$. Plotting the data
against the scaling variable $T/J(L)=T/L^\theta$, we obtain the best collapse
at $\theta=-0.45$, in agreement with the zero temperature
analysis presented above.

\begin{figure}
\epsfxsize=\linewidth
\epsfbox{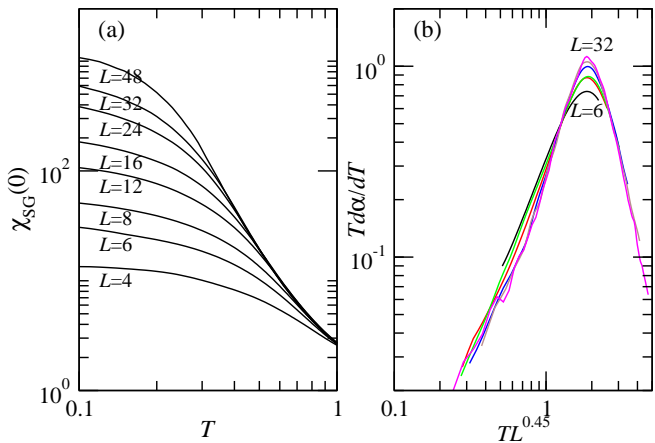}
\caption{(a) Spin-glass susceptibility versus temperature for various 
system sizes $L$. (b) Temperature derivative of the susceptibility
ratio at $|{\bf k}|=0$ and $2\pi/L$ versus $T/J(L)=TL^{-\theta}$ for six
sizes ranging from $L=6$ to 32. See text for details.
}
\label{chi_S}
\end{figure}

To summarize, the picture that emerges from our study of the 2D gauge glass
model is a ground state that supports gapless, fermion-like vortex-antivortex 
pair excitations. The energy scale of these excitations decays as a power-law 
of pair size --- a consequence of dielectric screening by pairs of smaller size.
The hierarchical organization of low energy vortices,
with diminishing energy scales on increasing length scales,
offers a new paradigm of zero-temperature criticality. 
This description is fully consistent with previous numerical work
as well as our low-temperature specific heat and spin-glass susceptibility 
data obtained under a multicanonical Monte Carlo sampling scheme.

The composite nature of the low energy vortex-antivortex pairs of large
size, with smaller and localized pairs providing screening of the long-ranged
Coulomb potential, suggests possible glassy dynamic behavior.
In a nonequilibrium context, the effective energy scales associated with 
long-distance vortex motion may be much higher than the equilibrium ones 
that require screening vortices to find their ideal positions.
Previous simulation work on the
relaxational dynamics of the gauge glass model indicates a rich
behavior with multiple size-dependent time scales.\cite{Hyman,NW04}
It would be interesting to re-analyze the data in light of the new insight
on the energetics of elementary vortex excitations.

We are grateful to T. K. Ng and S.-L. Wan for helpful discussions.
LH would like to thank Hao Li at UC San Francisco for hospitality, where
the work was completed.
Research is supported in part by the Research Grants Council of the 
Hong Kong SAR under grants HKBU 2061/00P and HKBU 2017/03P, and by the 
HKBU under grant FRG/01-02/II-65. Computations were carried out at HKBU's 
High Performance Cluster Computing Centre Supported by Dell and Intel.


\begin{thebibliography}{99}

\bibitem{Anderson} P. W. Anderson, {\it Basic Notions of Condensed Matter 
Physics}, 2nd Ed., Westview Press, Boulder, 1997.

\bibitem{FBJ} M. E. Fisher, M. N. Barber, and D. Jasnow, Phys. Rev. A 
{\bf 8}, 1111 (1973).

\bibitem{NK77} D. R. Nelson and J. M. Kosterlitz, 
Phys. Rev. Lett. {\bf 39}, 1201 (1977).

\bibitem{Dekker} C. Dekker, P. J. M. Wöltgens, R. H. Koch, B. W. Hussey 
and A. Gupta, Phys. Rev. Lett. {\bf 69}, 2717 (1992).

\bibitem{Fisher89} M. P. A. Fisher, Phys. Rev. Lett. {\bf 62}, 1415 (1989);
D. S. Fisher, M. P. A. Fisher, and D. A. Huse, Phys. Rev. B {\bf 43}, 130 (1991).

\bibitem{newrock}  R. S. Newrock, C. J. Lobb, U. Geigenm\"uller, and M.
Octavio, in {\it Solid State Physics}, Vol. 54, edited by F. Seitz {\it et al}.
(Academic Press, New York, 2000).

\bibitem{Hyman} R. A. Hyman, M. Wallin, M. P. A. Fisher, S. M. Girvin, 
and A. P. Young, Phys. Rev. B {\bf 51}, 15304 (1995).

\bibitem{Ebner} C. Ebner and D. Stroud, Phys. Rev. B {\bf 31}, 165 (1985).

\bibitem{FTY} M. P. A. Fisher, T. A. Tokuyasu and A. P. Young, 
Phys. Rev. Lett. {\bf 66}, 2931 (1991).

\bibitem{RY93} J. D. Reger and A. P. Young, J. Phys. A {\bf 26}, 1067 (1993).

\bibitem{Akino} N. Akino and J. M. Kosterlitz, Phys. Rev. B {\bf 66},
054536 (2002).

\bibitem{Katzgraber} H. G. Katzgraber, Phys. Rev. B {\bf 67}, 180402(R) (2003).

\bibitem{NW04} M. Nikolaou and M. Wallin, Phys. Rev. B {\bf 69}, 
184512 (2004).

\bibitem{CP99} M. Y. Choi and S. Y. Park, Phys. Rev. B {\bf 60}, 4070 (1999).

\bibitem{Kim}  B. J. Kim, Phys. Rev. B {\bf 62}, 644 (2000).

\bibitem{HMK} P. Holme, B. J. Kim, and P. Minnhagen, 
Phys. Rev. B {\bf 67}, 104510 (2003).

\bibitem{rubi83} M. Rubinstein, B. Shraiman, and D. R. Nelson,
Phys. Rev. B {\bf 27}, 1800 (1983).

\bibitem{nskl} T. Nattermann, S. Scheidl, S. E. Korshunov, and M. S. Li, 
J. Phys. I (France) {\bf 5}, 565 (1995); S. E. Korshunov and T. Nattermann,
Phys. Rev. B {\bf 53}, 2746 (1996); S. Scheidl, Phys. Rev. B {\bf 55}, 457 (1997).

\bibitem{tang96} L.-H. Tang, Phys. Rev. B {\bf 54}, 3350 (1996).

\bibitem{SE84} B. I. Shklovskii and A. L. Efros, 
{\it Electronic Properties of Doped Semiconductors} (Springer, Berlin, 1984).

\bibitem{note} Eq. (\ref{rho_R}) is expected to follow
from a more complete calculation which also yields a prescription for $c$. 
Such a strong-coupling theory lies outside our present scope.
Nevertheless, a case can be made for the finiteness of $c$ by noting
that pair states of negative energy in the specified size range
can not exceed a few in an area $R^2$.

\bibitem{berg} B. A. Berg and T. Neuhaus, Phys. Rev. Lett. {\bf 68}, 9 (1992).


\end{thebibliography}
\end{document}